\documentclass[11pt,a4paper]{scrartcl}

\usepackage{ILD}

\usepackage[symbol]{footmisc}
\usepackage{feynmf}
\usepackage{multirow}
\usepackage{textpos}

\title{Higgs self-coupling measurement at the International Linear Collider}

\ildproc{phys}{2023}{001} 

\date{\today}

\addauthor{Julie Munch Torndal}{\institute{1,2}}
\addauthor{Jenny List}{\institute{1}}

\addinstitute{1}{Deutsches Elektronen-Synchrotron DESY, Germany}
\addinstitute{2}{Universit\"at Hamburg, Department of Physics, Germany}

\abstract{The Higgs sector of particle physics is still largely uncovered, where establishing the Higgs mechanism is central to advance the field. The Higgs self-coupling is the key ingredient missing and an important puzzle piece for potentially uncovering new physics beyond the standard model. With the energy reach and precision reach of linear $e^+e^-$ colliders, the Higgs self-coupling can be measured directly and precisely enough that certain BSM scenarios can be evaluated. A new analysis of the capability to measure the Higgs self-coupling at the International Linear Collider (ILC) is ongoing and have identified aspects concerning the reconstruction tools which are expected to improve precision reach and are presented. This ongoing analysis intends to update the state-of-the-art projections for measuring the Higgs self-coupling at ILC which was previously evaluated at a centre-of-mass energy of 500 GeV. Additionally, the ongoing analysis intends to evaluate the choice of centre-of-mass energy and how it influences the reachable precision, as well as to consider how BSM effects might influence the reachable precision.\\\\
  
Talk presented at the International Workshop on Future Linear Colliders (LCWS 2023), 15-19 May 2023. C23-05-15.3.
}

\graphicspath{ {./logos/}{./figures/} }

\begin{document}

\titlepage

\section{Introduction}

The Higgs potential for the Higgs field, $h$, in the Standard Model (SM) after spontaneous symmetry breaking is defined as 
    \begin{equation}
        V(h)=\frac{1}{2}m_H^2h^2 + \lambda_3\nu h^3 + \frac{1}{4}\lambda_4h^4,
    \end{equation}
where $m_H$ is the Higgs mass, $\nu$ is the vacuum expectation value, $\lambda_3$ is the triple Higgs self-coupling and $\lambda_4$ is the quadratic Higgs self-coupling. In the SM, the Higgs self-couplings are predicted as
\begin{equation}
    \lambda_3^{SM}=\lambda_4^{SM}=\frac{m_H^2}{2\nu^2}.
\end{equation}
Having the measurement of the Higgs self-coupling, $\lambda$, would allow the shape of the Higgs potential to be determined, as well as to establish the Higgs mechanism experimentally from where we could determine how the Universe froze in the electroweak sector, giving mass to gauge bosons, fermions and the Higgs itself. Moreover, any deviations in $\lambda$ from $\lambda_3^{SM}$ would indicate new physics in the Higgs sector.  

\section{Measuring the Higgs self-coupling} \label{sec:lambdameasurement}
\begin{figure}[b]
\centering
\begin{subfigure}{.5\textwidth}
  \centering
  \includegraphics[height=4cm]{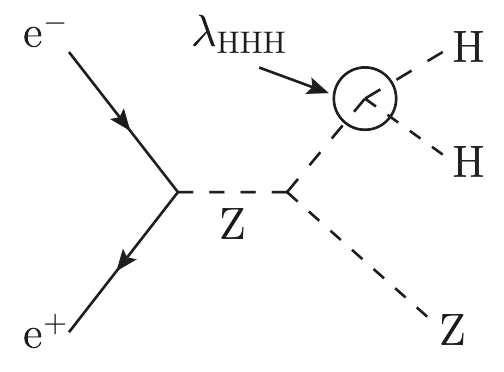}
  \caption{Di-Higgs strahlung}
  \label{fig:ZHH}
\end{subfigure}%
\begin{subfigure}{.5\textwidth}
  \centering
  \includegraphics[height=4cm]{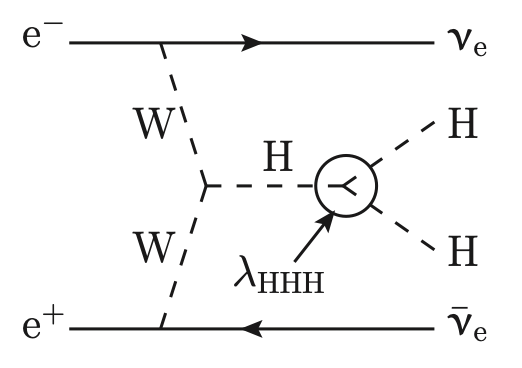}
  \caption{WW fusion}
  \label{fig:WWfusion}
\end{subfigure}
\caption{Main Higgs pair production modes containing the Higgs self-coupling.}
\label{fig:HHprod}
\end{figure}

\begin{figure}
\centering
\begin{subfigure}{\textwidth}
  \centering
  \includegraphics[height=3cm]{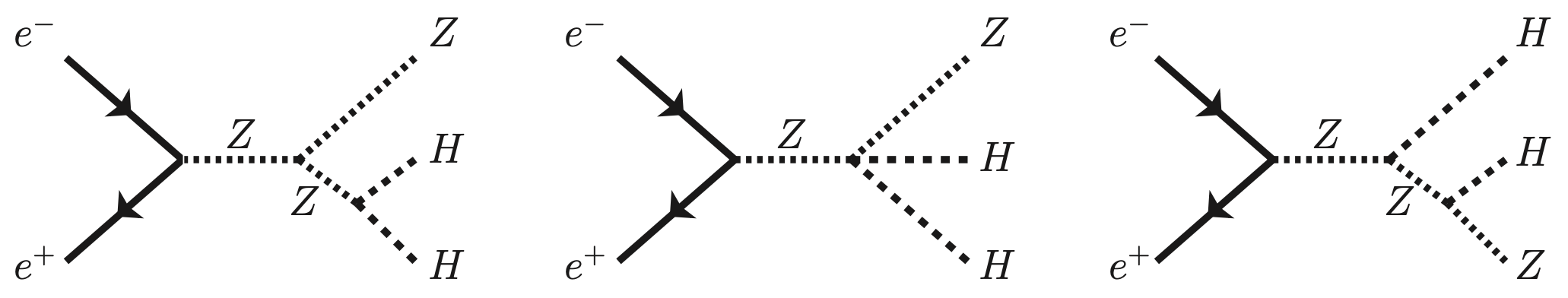}
  \caption{Di-Higgs strahlung interference diagrams}
  \label{fig:ZHHinterference}
\end{subfigure} \\
\begin{subfigure}{\textwidth}
  \centering
  \includegraphics[height=3.5cm]{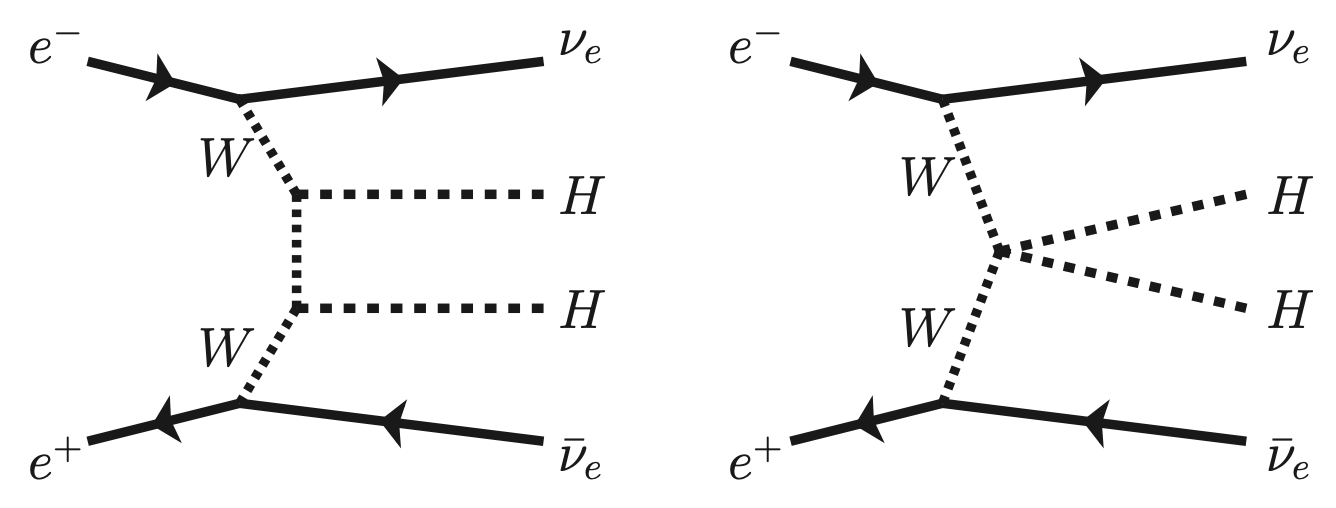}
  \caption{WW fusion interference diagrams}
  \label{fig:WWfusioninterference}
\end{subfigure}
\caption{Higgs pair production interference diagrams not containing the Higgs self-coupling.}
\label{fig:HHinterference}
\end{figure}

\begin{figure}
    \centering
    \includegraphics[width=0.8\textwidth]{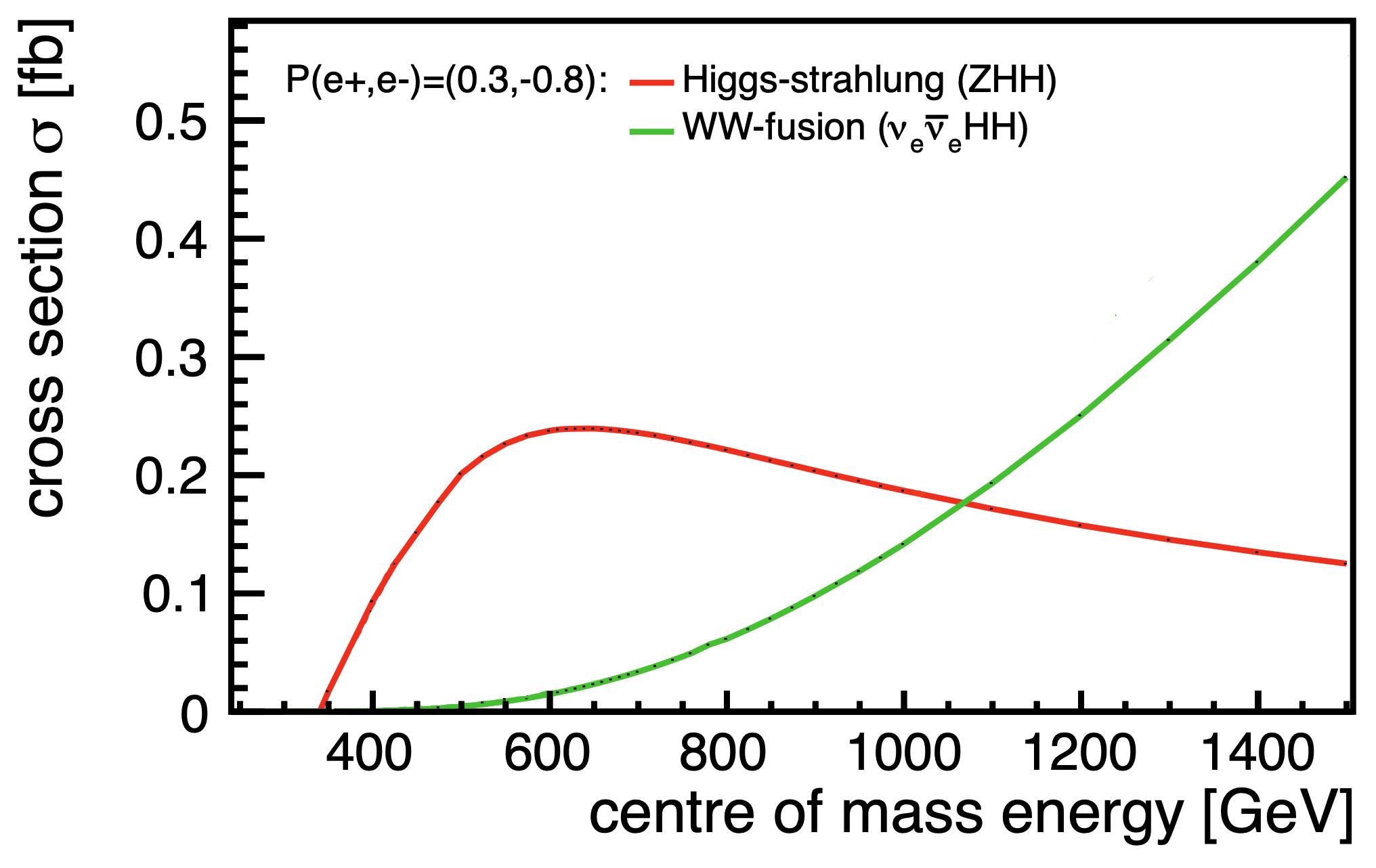}
    \caption{Cross sections for double Higgs production as a function of centre-of-mass energy (Modified from~\cite{Duerig:310520}).}
    \label{fig:hhxsec}
\end{figure}

The Higgs self-coupling can be measured either directly  or indirectly. Indirect access is provided through loop-order-corrections found from Effective Field Theory (EFT) fits using single Higgs measurements and running at two different centre-of-mass energies~\cite{Micco_2020}. Direct access is given through the measurement of double Higgs production. By measuring the cross section, the value of $\lambda$ can be inferred. This contribution focuses on the direct measurement of the Higgs self-coupling. At $e^+e^-$ colliders, there are two main production modes, di-Higgs strahlung and WW fusion. The Feynman diagrams containing the Higgs self-coupling for these two production modes are shown in Figure~\ref{fig:HHprod}. Figure~\ref{fig:HHinterference} shows the remaining Feynman diagrams not containing the Higgs self-coupling that still lead to the same final state. The diagrams for the $ZHH$ final state has constructive interference while the diagrams for the $\nu\nu HH$ final state has destructive interference.  Figure~\ref{fig:hhxsec} shows the cross sections for the processes as a function of the centre of mass energy. Di-Higgs strahlung is dominant below 1 TeV. Above 1 TeV, WW fusion dominates the di-Higgs production. The figure also shows that a future Higgs factory requires substantial energy to provide direct access to the Higgs self-coupling. The ILC physics programme has a planned running stage at 500 GeV (ILC500) with the possibility for upgrading to 1 TeV (ILC1000), hence at the ILC, Higgs pairs would be produced. This contribution focuses on the planned run at 500 GeV where the main production mode is $ZHH$ events which is a challenging measurement due to low statistics where only around 400 events are expected in total for all decay modes of the $Z$ and the Higgs bosons. \\

The state-of-the-art projections, that we have at ILC500, were performed 7-10 years ago in~\cite{Duerig:310520}. The analysis found that after the full ILC running scenario combining the channels with $HH\rightarrow bbbb$ and $HH\rightarrow bbWW$, the expected precision reach on the $ZHH$ cross section is 16.8\% which corresponds to an expected precision of 26.6\% on $\lambda_{SM}$. Combining the measurement at ILC500 with an additional running scenario at ILC1000, the precision could be further improved to 10\%. Hence, the discovery potential of the Higgs self-coupling at the ILC has been clearly demonstrated in the past already. However, the last analysis also identified several aspects mostly concerning the reconstruction tools that limits the precision reach. In the past decade, significant improvements concerning our reconstruction tools have been achieved which are expected to improve the expected precision reach on $\sigma_{ZHH}$ and hence also on $\lambda_{SM}$, making redoing the analysis of priority. \\

The first step of the analysis strategy concerning event reconstruction is removing the overlay from photons radiating of the initial beams that collide and produce low-$p_T$ hadrons. These hadrons need to be removed to uncover the underlying event. The analysis divides the $ZHH$ events into three channels; a lepton channel, a neutrino channel and a hadron channel referring to the decay of the $Z$-boson, and searches for $HH\rightarrow bbbb$ for all three channels. Once the overlay has been removed, an isolated lepton tagging is applied to identify leptons for selection or rejection depending on the channel and a lepton pairing is done in the case of the lepton channel. The remaining event is clustered into four or six jets and flavor tagging is applied to search for the many $b$-jets. A preselection is applied for each channel before kinematic fitting is performed to help separate $ZHH$ events from $ZZH$ events which are the main irreducible background, and an event selection can be applied. Some of the aspects concerning reconstruction, that are limiting the reachable precision, are highlighted below. \\

\begin{figure}
    \centering
    \includegraphics[width=0.8\textwidth]{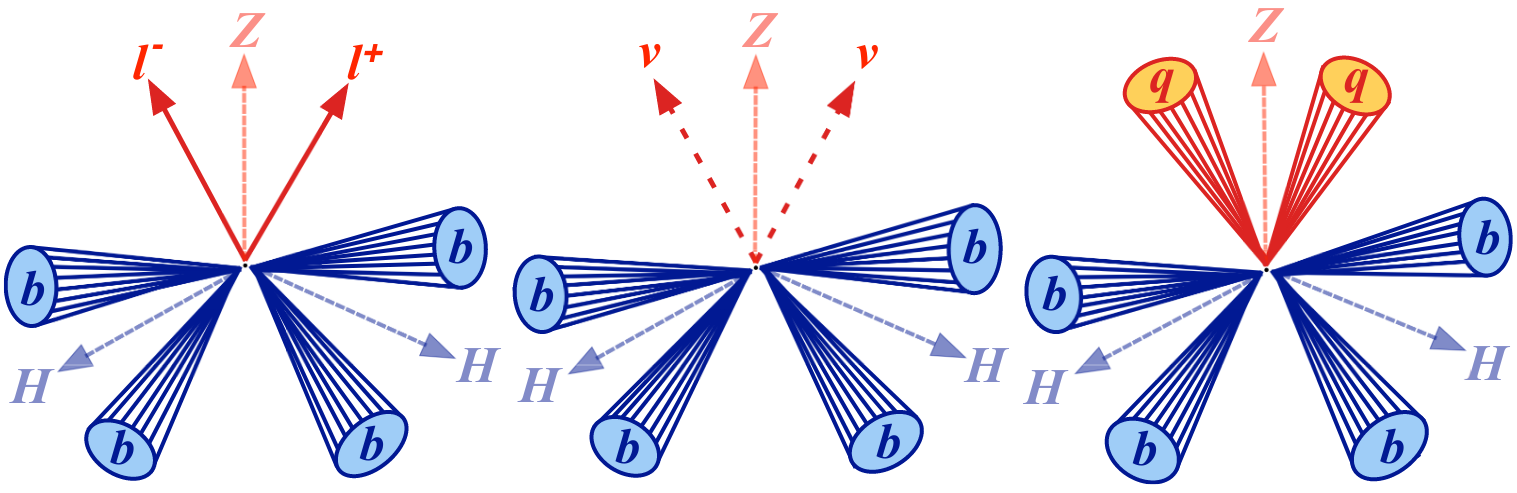}
    \caption{Signature of the three search channels with the Higgs pair decaying to four $b$-jets and $Z$ decaying to a lepton pair in the lepton channel, missing energy from the neutrino pair in the neutrino channel, and additionally two jets in the hadron channel. Image from~\cite{Duerig:310520}.}
    \label{fig:signature}
\end{figure}

\begin{description}
\item[Overlay removal] Since the last analysis, a better modelling of the overlay has been achieved which would lead to more realistic results. Additionally, fewer overlay events are expected. In the last analysis, 1.7 events were expected on average while now the number has been reduced to 1.05 events on average. The standard overlay removal strategy clusters the low-$p_T$ hadrons into very forward beam jets that are removed to uncover the original events. However, with $ZHH$ events, the jets are not very boosted and they tend to overlap causing misclustering which might complicate the overlay removal. More detailed studies are needed to determine whether a more advanced removal strategy is needed. 
\item[Isolated lepton tagging] The past analysis searched only for electrons and muons as the MVA based approach~\cite{DuerigTian_presentation} was optimised for identifying those particles and did not contain a dedicated search for tau leptons. For comparison, if the same efficiency for identifying electrons and muons could be reached for identifying tau leptons, that would lead to an 8\% relative improvement in the precision~\cite{Duerig:310520}. Since the last analysis, a method has been developed that reconstructs taus using impact parameters~\cite{Jeans_2016}. The method requires an accurate $\tau$ vertex and precise measurements of the decay products which we would have with the ILD. 
\item[Jet clustering] One aspect, that limits the precision significantly, is that of jet clustering. The fact that jets in the $ZHH$ events tend to overlap causes misclustering and pairing issues. Having perfect jet clustering, where those issues are non-existent, would lead to a 40\% relative improvement in the precision~\cite{Duerig:310520}. More details can be found in section~\ref{sec:jetclustering}. Developments in new methods for jet clustering remains as wide open questions, although historically, each of the shifts to the $e^+e^-$ collider era of LEP and the $pp$ collider era of LHC lead to huge paradigm shifts in the field of jet clustering. With the future shift back to an $e^+e^-$ collider era with a future Higgs factory, one could expect one more such paradigm shift. 
\item[Flavor tagging] Sticking with jets clustered using our tried-and-tested algorithms, flavor tagging can be applied. This is an aspect that has seen improvements. The past analysis quoted that just a 5\% relative improvement in the efficiency for identifying $b$-jets would lead to an 11\% relative improvement in the precision~\cite{Duerig:310520}. This is something that has been achieved. Section~\ref{sec:flavortagging} expands further on the aspect of flavor tagging. 
\item[Error parametrisation in kinematic fitting] The last aspect that will be highlighted in this contribution is that of error parametrisation for kinematic fitting. The mass resolution of jets depend on the jet energy resolution. Since the last analysis, a new tool called ErrorFlow~\cite{radkhorrami2021kinematic} has been developed. Section~\ref{sec:kinfitting} expands further on ErrorFlow and its effect on the performance of kinematic fitting. 
\end{description}

Improvements in the reconstruction tools mentioned above has the potential to bring the sensitivity to better than 20\% at ILC500~\cite{aryshev2023international}. 

\section{Jet clustering} \label{sec:jetclustering}
\begin{figure}
\centering
\begin{subfigure}{.5\textwidth}
  \centering
  \includegraphics[width=\textwidth]{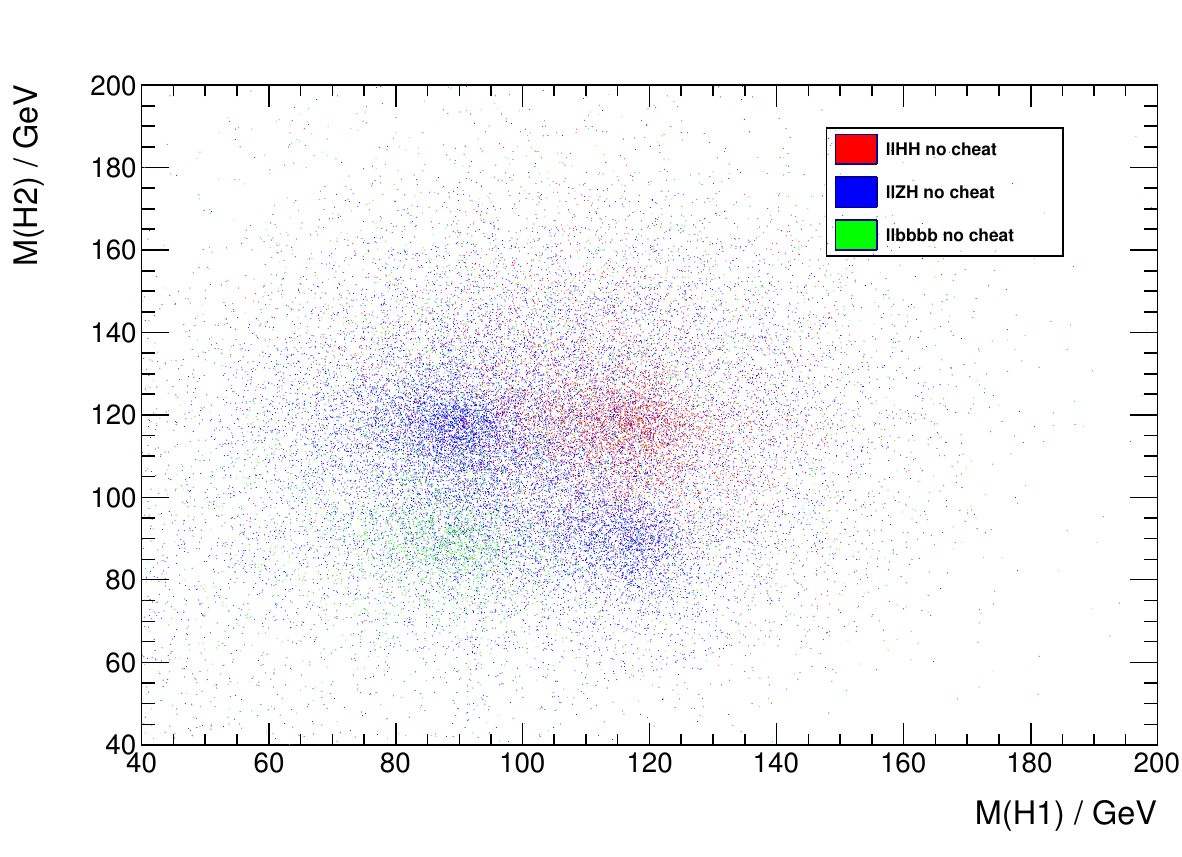}
  \caption{Durham jet clustering}
  \label{fig:durhamjetclustering}
\end{subfigure}%
\begin{subfigure}{.5\textwidth}
  \centering
  \includegraphics[width=\textwidth]{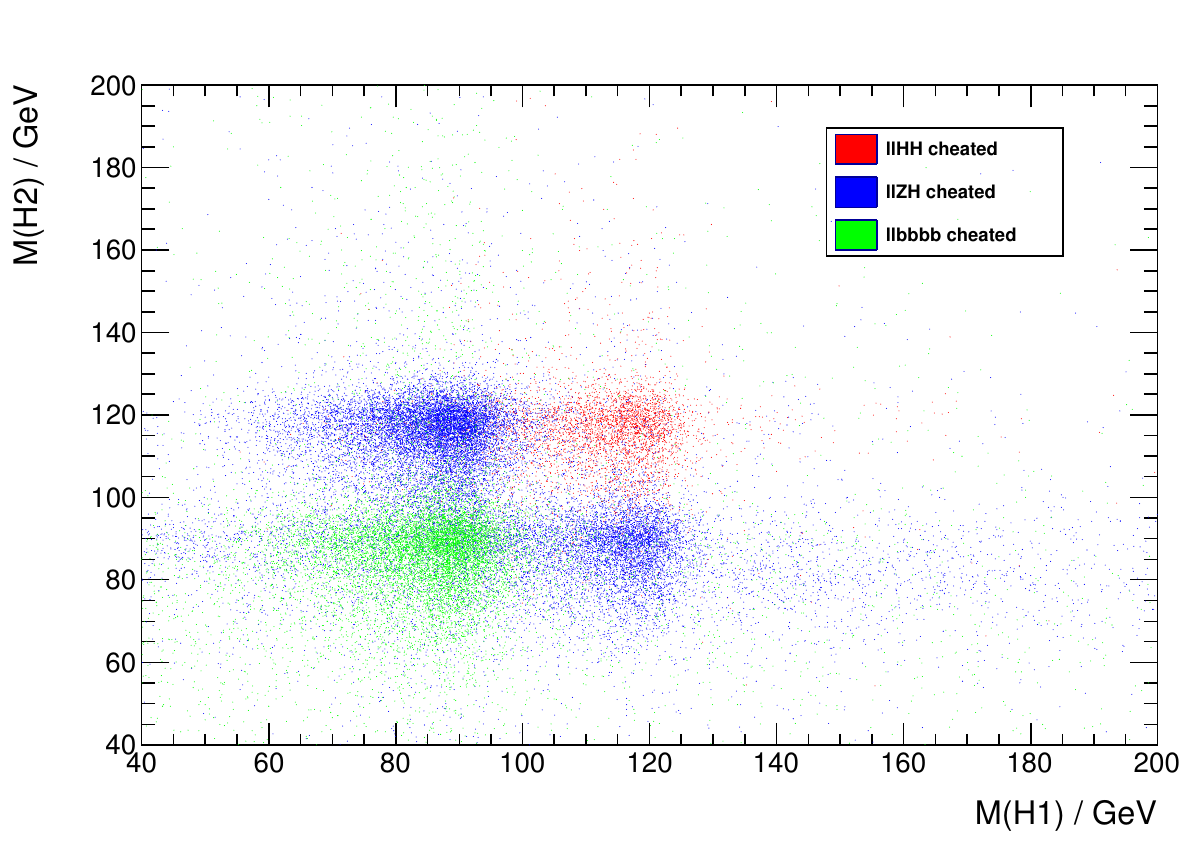}
  \caption{Perfect jet clustering}
  \label{fig:perfectjetclustering}
\end{subfigure}
\caption{Dijet masses from a signature with 2 leptons and 4 jets that are paired into the two dijet-systems. Signal events from $ZHH$ are concentrated in the region where both dijets have a mass close to the Higss mass. Similarly, background events from $ZZH$ and $ZZZ$ are concentrated regions with dijet masses corresponding to the initial bosons. Reconstructed jets clustered with the Durham algorithm (a) show large overlap compared to jets from cheated clustering to achieve perfect jet clustering (b)~\cite{junpingplots}.}
\label{fig:jetclustering}
\end{figure}

\begin{figure}
    \centering
    \includegraphics[width=0.8\textwidth]{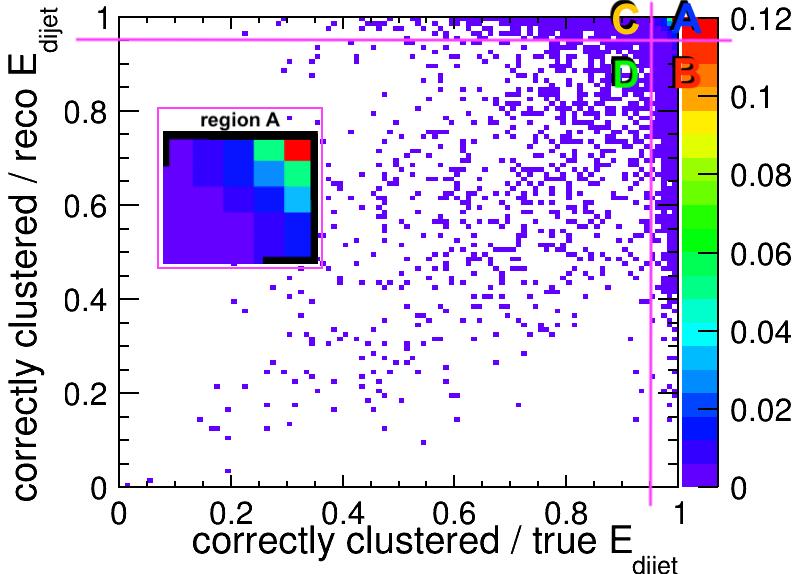}
    \caption{Energy fraction of correctly clustered PFOs in true jets versus energy fraction of correctly clustered PFOs in reconstructed jets. Correctly clustered PFOs are defined as PFOs that are in both a reconstructed jet and the corresponding true jet that has been matched as the true jet closest in angular space. }
    \label{fig:Misclustering}
\end{figure}

\begin{figure}
    \centering
    \includegraphics[width=0.8\textwidth]{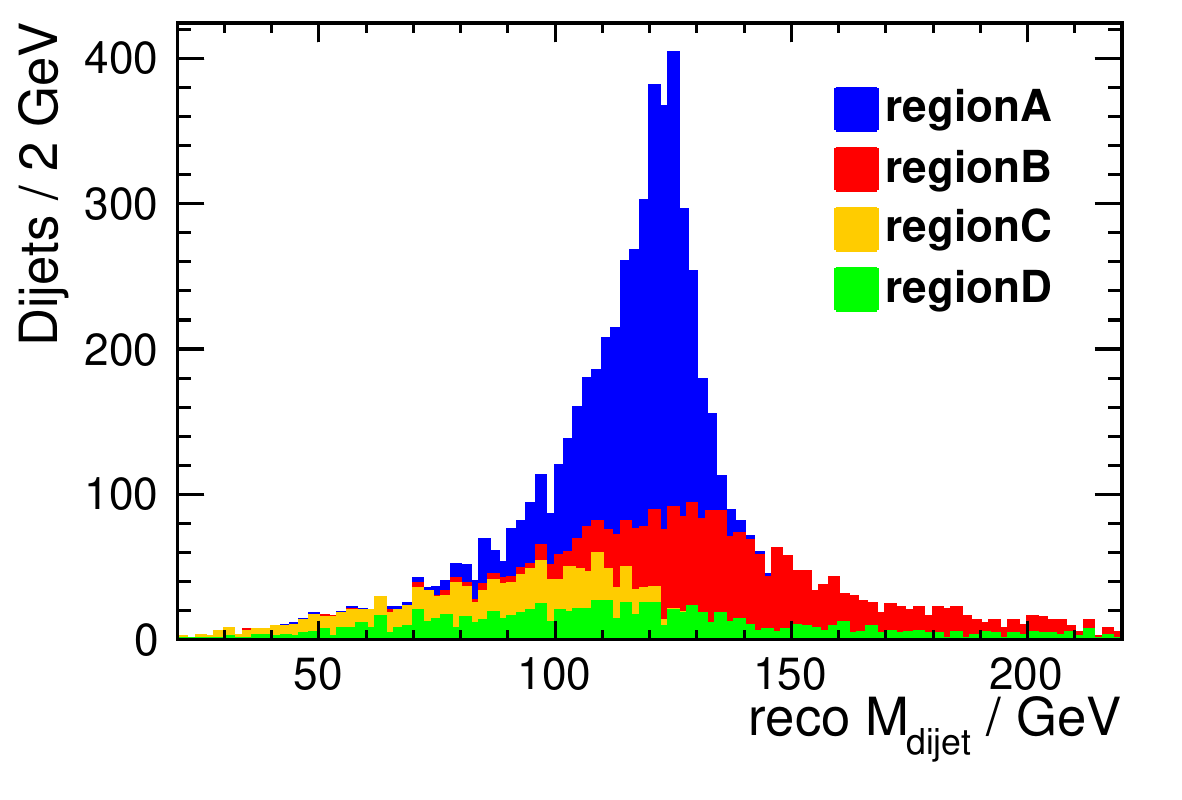}
    \caption{Dijet mass distributions for $ZHH$ events stacked for each four regions A, B, C, and D showing how misclustering worsens the resolutions of the dijet masses. The definition of the four regions can be seen from Figure~\ref{fig:Misclustering}.} 
    \label{fig:misclusteringdijetsmasses}
\end{figure}

The two plots in Figure~\ref{fig:jetclustering} show the dijet masses in the case of using the Durham algorithm to cluster the jets~(\ref{fig:durhamjetclustering}) and the case of perfect jet clustering~(\ref{fig:perfectjetclustering}). Comparing the two plots shows that jet finding ambiguities degrades the mass resolutions which reduces the separation of signal and background. Hence, the sensitivity to the Higgs self-coupling is reduced by almost a factor $\sim2$~\cite{Duerig:310520}. Using a tool called TrueJet~\cite{Berggren:416718}, one can cluster together the reconstructed particle flow objects (PFOs) using the Monte Carlo history for the corresponding true particles which ensures that dijets are clustered together according to the truth. By matching together the true jets and the reconstructed jets in angular space, the PFOs appearing in both a true and corresponding reconstructed jet are assumed to be correctly clustered. Figure~\ref{fig:Misclustering} shows the fraction of energy in a true dijet that is correctly clustered versus the fraction of energy in a reconstructed dijet that is correctly clustered.  The plot is divided into four regions; A, B, C, and D. Most of the events end up in region A where the reconstructed dijets have good agreement with the true dijets pointing to no misclustering issues. But there are also large populations in regions B, C, and D. Figure~\ref{fig:misclusteringdijetsmasses} shows the mass distribution of the dijets for each the four regions stacked. Dijets in region A accounts for majority of the events in the centre of the distributions while populations in regions B, C, and D account for the tails. The issue of misclustering is something that is still being investigated but the plots shown here showcase how badly advanced jet clustering is needed to address the misclustering regions B, C, and D. \\

\section{Flavor tagging} \label{sec:flavortagging}
\begin{figure}
    \centering
    \includegraphics[width=0.8\textwidth]{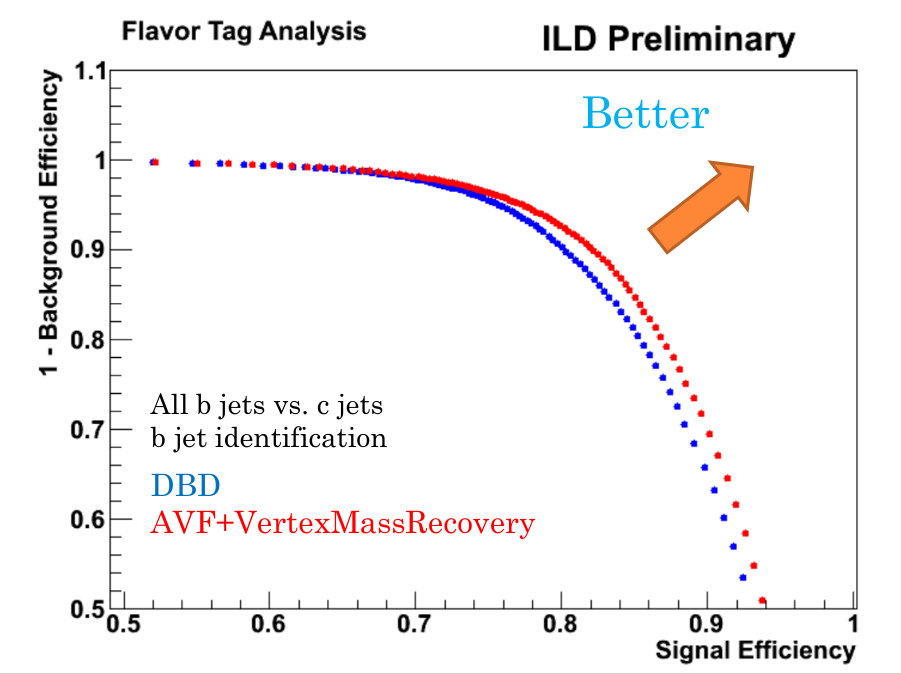}
    \caption{$b/c$ separation using LFCIplus. The ROC curves compare the baseline design of the algorithm with an update from 2017. More details can be found in \cite{Taikan_presentation}.}
    \label{fig:flavortagging}
\end{figure}

As mentioned in section~\ref{sec:lambdameasurement}, identifying $b$-jets has improved which can be seen from Figure~\ref{fig:flavortagging} where the $b/c$ separation is shown from the ROC curves comparing the baseline design with an update in LCFIplus~\cite{Suehara_2016} from 2017~\cite{Taikan_presentation}. As an example, at 80\% signal efficiency, 90\% of the $c$-jets could be rejected in the baseline design while 95\% of the $c$-jets could be rejected with the update. This corresponds to doubling the rejection factor from 10 to 20. This improved $b$-tagging leads to larger signal efficiencies in the preselection when comparing to the past analysis. The cutflow for the preselection in the neutrino channel is shown in Table~\ref{tab:cutflowtable} and shows a 74\% relative improvements after the cut on $b$-likelihoodness of the third jet when comparing the signal efficiencies between the ongoing and past analysis. 

\begin{table}[]
    \centering
    \begin{tabular}{c}
         \begin{minipage}{\textwidth}
      \includegraphics[width=\linewidth, height=60mm]{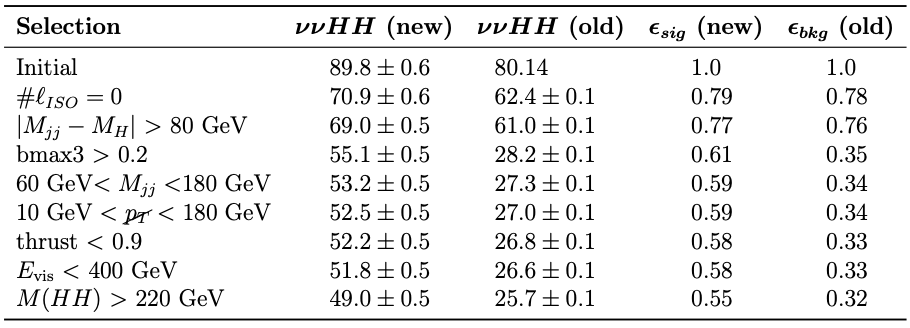}
    \end{minipage}
    \end{tabular}
    \caption{Cutflow in neutrino channel. The $bmax3$ is defined as the third largest $b$-tag value of the four jets. The details of the preselection can be found in~\cite{Duerig:310520}.}
    \label{tab:cutflowtable}
\end{table}

\section{Kinematic fitting} \label{sec:kinfitting}
\begin{figure}
    \centering
    \includegraphics[width=0.8\textwidth]{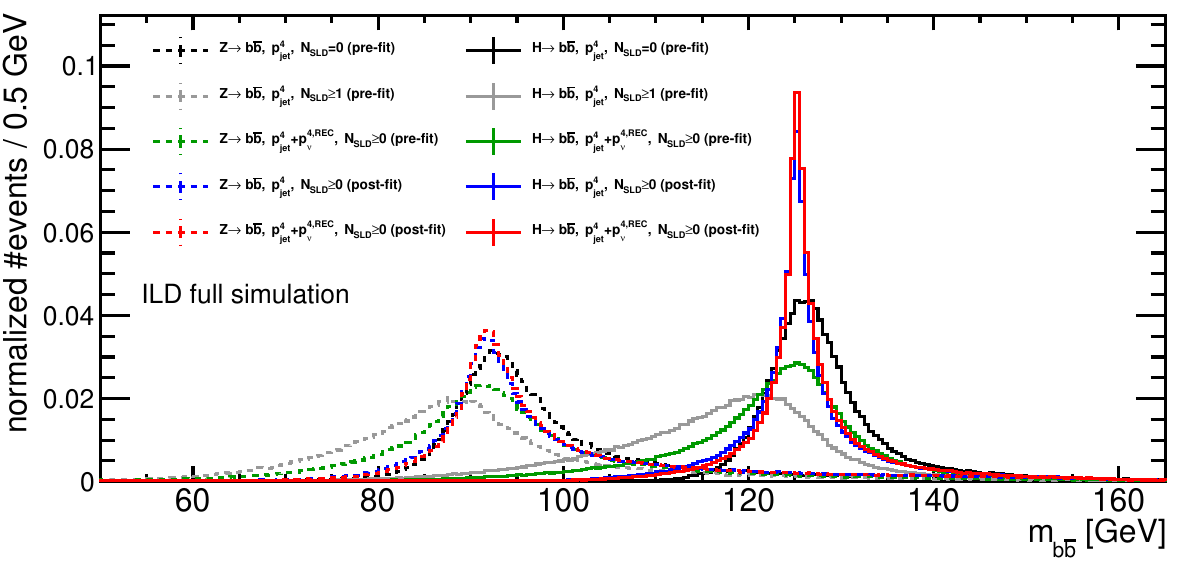}
    \caption{$Z/H\rightarrow b\Bar{b}$ mass distributions showing the effect of kinematic fitting using ErrorFlow and neutrino correction~\cite{yasserplot}. The grey, black, and green distributions shows the $b\Bar{b}$ mass distributions when there are 0 SLDs in the jets, at least 1 SLD in the jets and the combined distribution, respectively. The blue and red distributions show the the $b\Bar{b}$ mass distributions after a kinemaic fit excluding and including the neutrino correction, respectively.}
    \label{fig:errorflow}
\end{figure}

With kinematic fitting, measurements are corrected by imposing a set of constraints on them. This method can be used to improve the kinematics, e.g. mass resolution, perform hypothesis testing or simply to apply jet-pairing. The $\chi^2$-function that needs to be minimised is \\
\begin{equation}
    L(y)=\Delta y^T \mathbf{V}(y)^{-1} \Delta y+2\sum_{k=1}^m \lambda_k f_k(a,y),
\end{equation} \\
where $y$ is the  set of measured parameters, $a$ is the set of unmeasured parameters, $\Delta y$ are  corrections to $y$,
$\mathbf{V}(y)$ is the covariance matrix for $y$, $f_k$ is the set of constraints expressing the fit model, and $\lambda_k$ are the Lagrange multipliers. The covariance matrix is crucial for a well-performing fit. ErrorFlow is a tool that calculates the error parametrisation of the covariance matrix by parameterising several sources of uncertainties for each jet individually. So far, ErrorFlow accounts for detector resolution, particle confusion in the Pandora Particle Flow algorithm~\cite{marshall2013pandora}, as well as corrects for unmeasured neutrinos in the jets from semi-leptonic decays (SLDs) within the heavy $b$-jets. Figure~\ref{fig:errorflow} shows how the mass resolution improves with the kinematic fit. The green distributions show the $b\Bar{b}$ mass distributions from either a $Z$ or Higgs decay prior to any fitting. The grey and black distributions shows the masses for jets either containing or not containing any SLDs respectively. The resolution of the mass distributions are reduced by the presence of SLDs. The red and blue distributions are post-fit with and without the neutrino correction. While the kinematic fit on its own already improves the mass resolution significantly, including the neutrino correction improves it further. The plot also illustrates how kinematic fitting can be used to improve the separation of $Z$ and Higgs bosons. \\

\begin{figure}
\centering
\begin{subfigure}{.5\textwidth}
  \centering
  \includegraphics[width=0.95\textwidth]{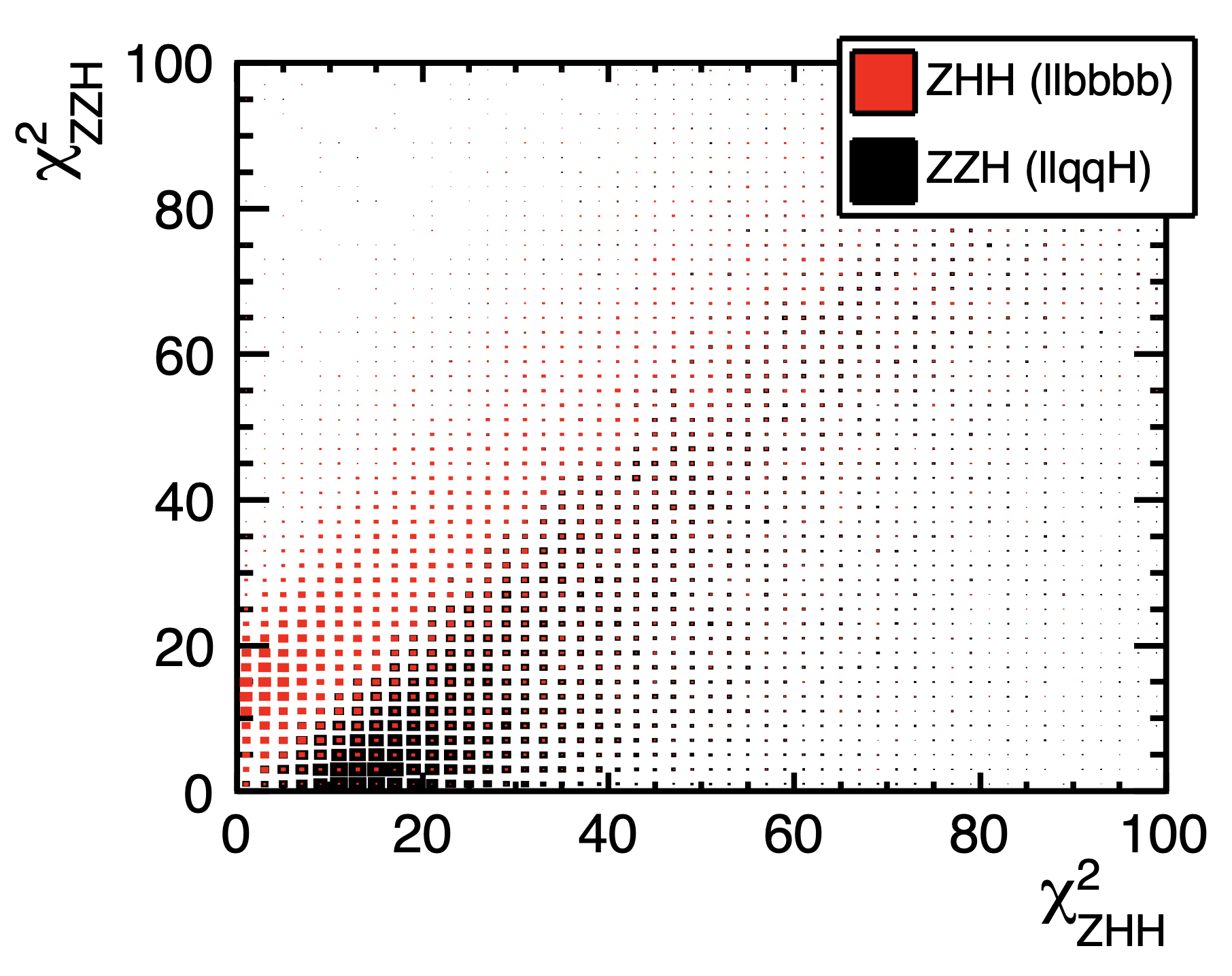}
  \caption{From~\cite{Duerig:310520} prior to ErrorFlow}
  \label{fig:kinfitold}
\end{subfigure}%
\begin{subfigure}{.5\textwidth}
  \centering
  \includegraphics[width=\textwidth]{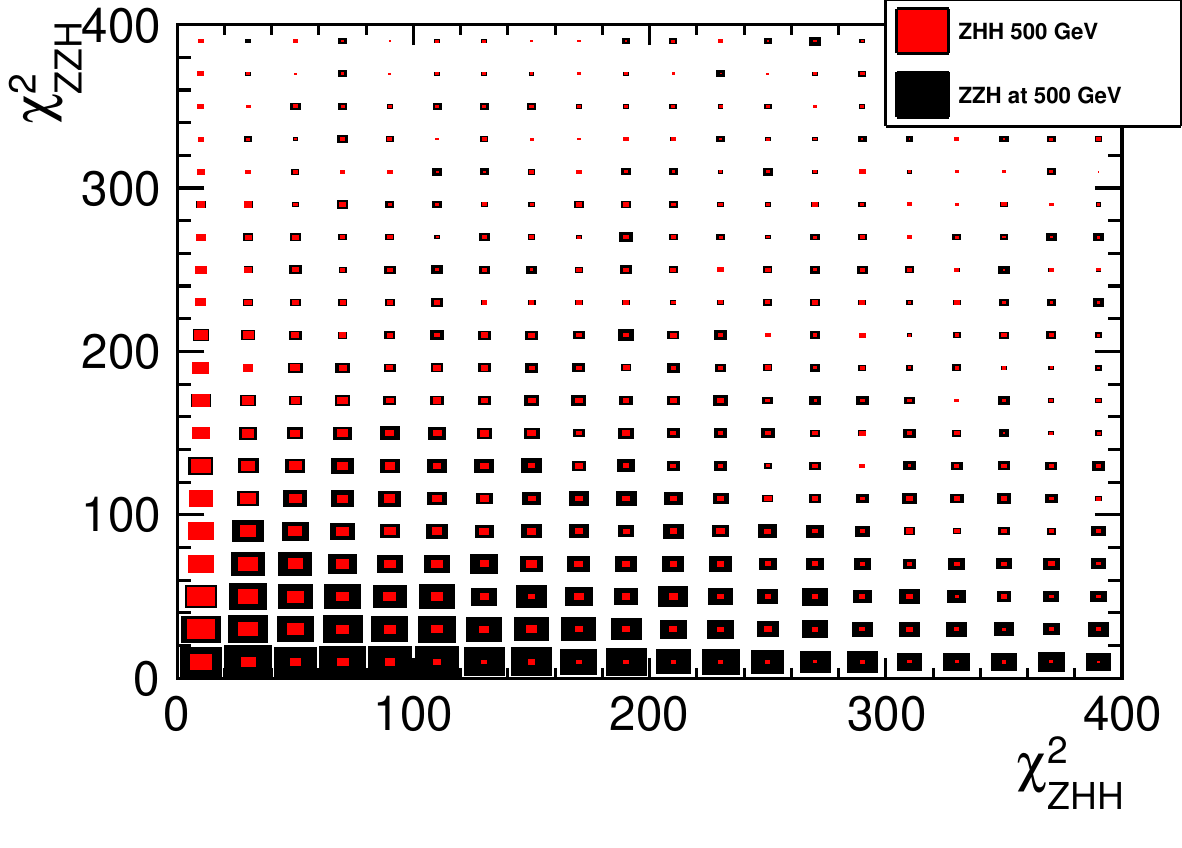}
  \caption{Using ErrorFlow}
  \label{fig:kinfitnew}
\end{subfigure}
\caption{Hypothesis testing using kinematic fitting comparing the previous analysis with the ongoing analysis for measuring the Higgs self-coupling at ILC. The distributions of $\chi^2_{ZHH}$ versus $\chi^2_{ZZH}$ for signal ($ZHH$) and background ($ZZH$) events have qualitatively changed dramatically between the two analyses. }
\label{fig:kinfit}
\end{figure}

In the context of the $ZHH$ analysis, kinematic fitting can also be utilised for hypothesis testing. Figure~\ref{fig:durhamjetclustering} showed a large overlap between signal and background events with one $Z$ giving a signature of two leptons and the remaining two bosons ($HH$, $ZH$, and $ZZ$) giving a signature of four jets. A $\chi^2$ for the $ZHH$ hypothesis and a $\chi^2$ for the $ZZH$ hypothesis can be calculated for each event. Both hypotheses imposes four constraints for assuming energy and momentum conservation and two mass constraints for each of the dijets where the $ZHH$ hypothesis assumes both dijets to fit with the Higgs mass and the $ZZH$ assumes one dijet to fit with the Higgs mass and the other dijet to fit with the $Z$ mass. Figure~\ref{fig:kinfit} illustrates the $\chi^2_{ZHH}$ versus $\chi^2_{ZZH}$ for signal and background events from the previous analysis~\ref{fig:kinfitold} as well for the ongoing analysis~\ref{fig:kinfitnew}. The previous analysis already showed good separation without ErrorFlow, especially in the region where $\chi^2$-values are low. With ErrorFlow, there is a significant qualitative difference where the majority of the $\chi^2$-values are pushed further from each other along the edges. However, there is still a large overlap between the $ZHH$ and $ZZH$ events which needs to be understood. Furthermore the ongoing analysis has an open question regarding the choice of centre-of-mass energy and would like to answer if and how the performance in flavor tagging and kinematic separation improves when increasing the centre-of-mass energy to 550 GeV and 600 GeV. 

\section{Precision on the Higgs self-coupling}
Table~\ref{tab:precision} shows how the precision reach on the Higgs self-coupling at ILC500 compares to other future colliders. ILC500 is competitive with CLIC that would also have direct access to the Higgs self-coupling, while FCC-ee only has indirect access with its physics programme planned at lower centre-of-mass energies. FCC-hh is projected to have the best precision reach on directly measuring the Higgs self-coupling, however, with the time-scales of these future colliders that measurement is a few decades further away. Additionally, the numbers for FCC-hh are based on fast simulations as opposed to the full simulation studies done for ILC. \\

\begin{figure}
\centering
\begin{subfigure}{.5\textwidth}
  \centering
  \includegraphics[width=\textwidth]{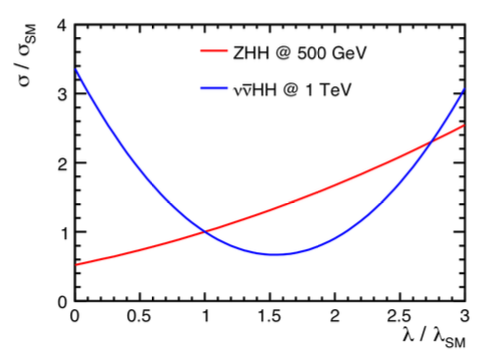}
  \caption{At ILC~\cite{tian_presentation}}
  \label{fig:xsecILC}
\end{subfigure}%
\begin{subfigure}{.5\textwidth}
  \centering
  \includegraphics[width=\textwidth]{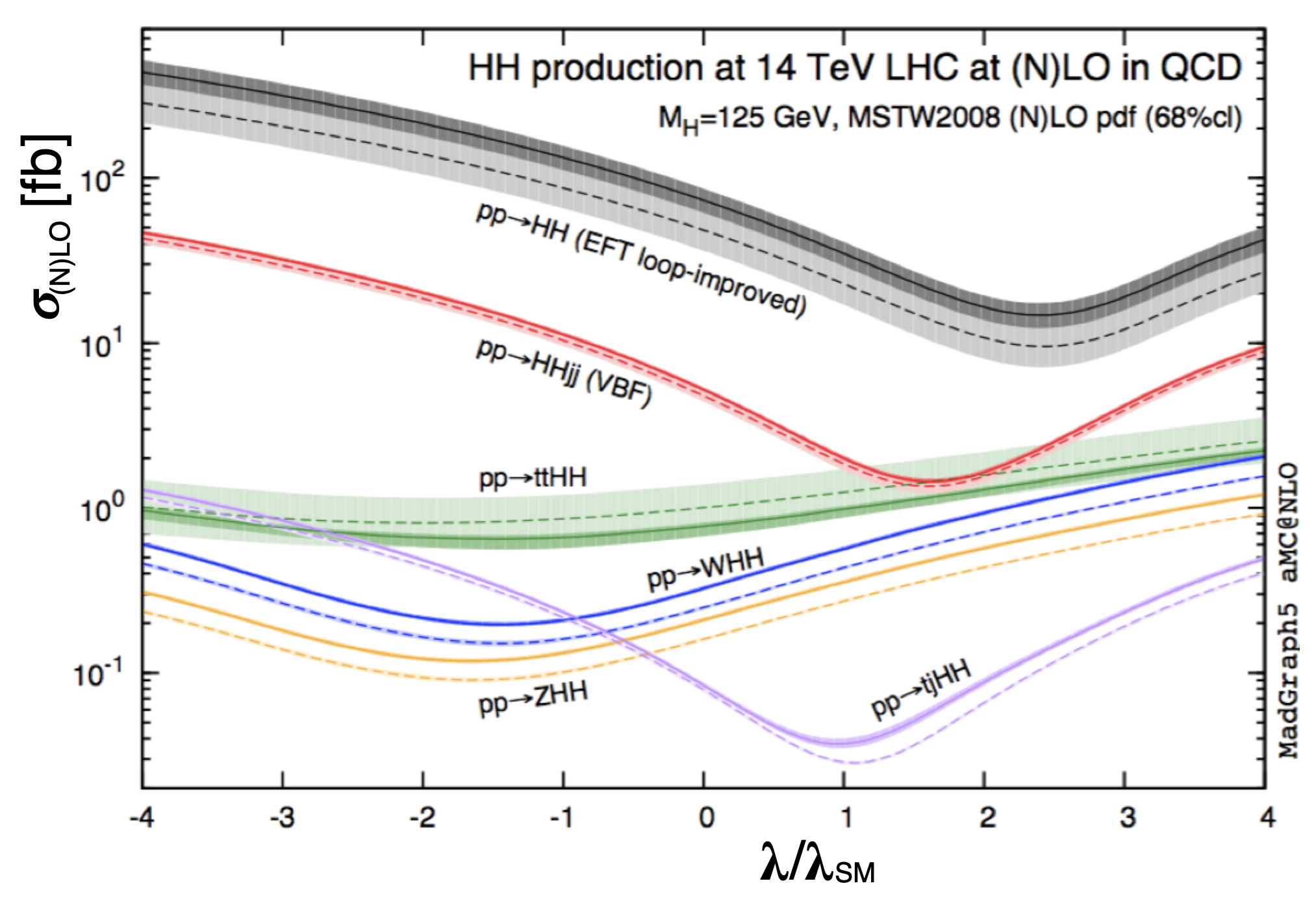}
  \caption{At LHC~\cite{Frederix_2014}}
  \label{fig:xsecLHC}
\end{subfigure}
\caption{Cross sections as a function of $\lambda$. The cross sections for $ZHH$ and $\nu\nu HH$ show a complementarity between the two production modes at $e^+e^-$ colliders, that is not seen to the same extent at $pp$ colliders.}
\label{fig:xsecforlambda}
\end{figure}

\begin{figure}
    \centering
    \includegraphics[width=0.8\textwidth]{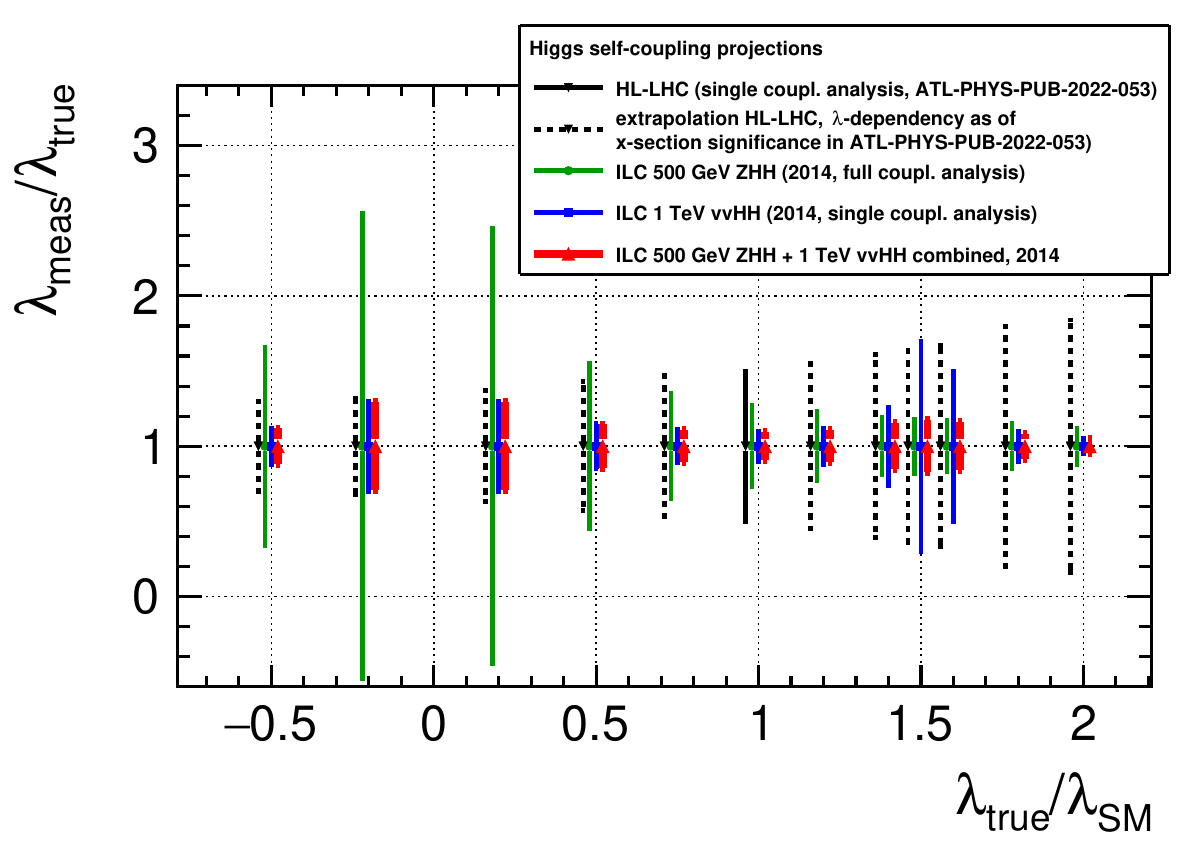}
    \caption{Comparison of various state-of-the-art projections for the Higgs self-coupling measurements at HL-LHC and ILC. The HL-LHC projections are based on~\cite{ATL-PHYS-PUB-2022-053}, assuming that the self-coupling measurement precision will change with the actual value of $\lambda$ similarly to the cross-section significance. The ILC projections are based on~\cite{Duerig:310520,LC-REP-2013-025} from 2014, and significant improvements will be expected from the ongoing re-analysis described in these proceedings. Note that only for the extraction of $\lambda$ from ZHH at 500GeV the shown values correspond to a global SMEFT extraction~\cite{Barklow:2017awn}, while all other projections are single-parameter fits.}
    \label{fig:precisionforlambda}
\end{figure}

\begin{table}[]
    \centering
    \begin{tabular}{ccc}
    \toprule
        collider & indirect-$h$ & direct-$hh$ \\ \midrule
        HL-LHC & 100-200\% & 50\% \\ \midrule
        ILC250 & - & - \\
        ILC500 & 58\% & 20\%$^*$ \\
        ILC 500/1000 & 52\% & 10\%$^*$ \\
        CLIC380 & - & - \\
        CLIC1500 & - & 36\% \\
        CLIC 1500/3000 & - & 9\% \\
        FCC-ee 240 & - & - \\
        FCC-ee 240/365 & 44\% & - \\
        FCC-ee (4 IPs) & 27\% & - \\
        FCC-hh & - & 3.4-7.8\% \\ \bottomrule
    \end{tabular}
    \caption{Precision reach of future colliders~\cite{Micco_2020},~\cite{narain2023future}. $^*$ The better than 20\% precision for ILC500 is projected under the assumption that the improvements seen in our reconstruction tools are propagated to a new analysis. Combining the measurement at ILC500 with an additional running scenatio at ILC1000 might also lead to a better than 10\% precision~\cite{aryshev2023international}.}
    \label{tab:precision}
\end{table}
It's important to keep in mind that the numbers listed in Table~\ref{tab:precision} are only valid when the value of $\lambda$ corresponds to $\lambda_{SM}$. The Higgs self-coupling measurement is special because the precision is highly dependent on its own value. Figure~\ref{fig:xsecforlambda} shows the cross sections for various modes of di-Higgs production at $e^+e^-$ and $pp$ colliders and how they depend on the value of $\lambda$. At $e^+e^-$ colliders, the two channels, di-Higgs strahlung and WW fusion, provide complementary information. $ZHH$ gives stronger constraints when $\lambda/\lambda_{SM}>1$ and $\nu\nu HH$ gives stronger constraints when $\lambda/\lambda_{SM}<1$. At LHC the constraints are stronger in the case of $\lambda/\lambda_{SM}<1$. The precision reach for $\lambda$ can be extrapolated from the precision reach on $\lambda_{SM}$ which is seen in Figure~\ref{fig:precisionforlambda}. The black bars show the expected precision reach for HL-LHC where the precision improves for lower values of $\lambda$. The green bars are the expected precision reach for $ZHH$ only and the blue bars are the expected precision reach for $\nu\nu HH$ only at ILC. $ZHH$ has the best precision reach for large values of $\lambda$ while $\nu\nu HH$ has the best precision reach for low values of $\lambda$. The expected precision reach when combining $ZHH$ and $\nu\nu HH$ events can be seen from the red error bars. By having both processes, at least 10-15\% precision can be ensured for \textit{any} value of $\lambda$. 

\section{Conclusion}
The discovery potential of the Higgs self-coupling at the ILC has been clearly demonstrated in the past. Since the state-of-the-art projections were performed, there has been significant improvements in the reconstruction tools which are expected to lead to an improvement of the sensitivity to better than 20\% at ILC500. An update to the state-of-the-art projections is underway to propagate the improvements in reconstruction tools to a new analysis on measuring the Higgs self-coupling at ILC through the measurement of $ZHH$ events and to answer how much better than 20\% we could do. This answer is important for shaping the landscape of future colliders. 

The improvements in reconstruction tools discussed in this contribution concern overlay, reconstruction of leptons, jet clustering, flavor tagging, and error parametrisation for kinematic fitting. Concerning overlay, a better and more realistic modelling of the overlay has been achieved and the number of overlay events that need to be removed has been reduced. However, more detailed studies are needed to determine whether an advanced removal strategy needs to be developed. Another strategy that has been developed concerns tau lepton reconstruction. The previous analysis on measuring the Higgs self-coupling at ILC was optimised for searching for electrons and muons but lacked a dedicated search for tau leptons which the ongoing analysis intends to include. Concerning jet reconstruction, this contributions highlights the need for advanced jet clustering. We discussed how misclustering in jets causes poor mass resolutions of the dijets limiting the separation power of signal and background. Another limiting factor pointed out in the previous analysis is that of flavor tagging. With the signature of 4 $b$-jets, $b$-tagging is crucial. Since the past analysis was performed, the flavor tagging in LCFIPlus has improved and the improvements have been propagated to the analysis showing better signal efficiencies. The last reconstruction tool presented in this contribution is that of ErrorFlow which parameteriseres several sources of uncertainties of the covariance matrices for individual jets. With ErrorFlow, a dramatic change has been observed for the $\chi^2$ distributions obtained from kinematic fitting. Some details still need to be understood. Additionally, this contribution has also opened up a question of reconsidered choice of centre-of-mass energy by looking into how the performance in flavor tagging and kinematic separation might improve at 550 GeV or 600 GeV. The final part of this contribution considered how BSM effect could influence the reachable precision on $\lambda$ and showed how the complementarity of ILC500 and ILC1000 is important to ensure at least 10-15\% precision for \textit{any} value of $\lambda$.

\section*{Acknowledgements}
We would like to thank the LCC generator working group and the ILD software working group for providing the simulation and reconstruction tools and producing the Monte Carlo samples used in this study.
This work has benefited from computing services provided by the ILC Virtual Organization, supported by the national resource providers of the EGI Federation and the Open Science GRID.

\printbibliography{}
\end{document}